\documentstyle[12pt]{article}

\batchmode
  \newfont{\smallfrakfont}{eufm8}
  \newfont{\smallbbbfont}{msbm8}
\errorstopmode
\newif\iffrakf\frakftrue
\ifx\smallfrakfont\nullfont\frakffalse\fi
\newif\ifamsf\amsftrue
\ifx\smallbbbfont\nullfont\amsffalse\fi
\iffrakf
  \newfont{\frakfont}{eufm10 scaled\magstep1}  
  \newfont{\tinyfrakfont}{eufm6}
\fi
\ifamsf
  \newfont{\bbbfont}{msbm10 scaled\magstep1}  
  \newfont{\tinybbbfont}{msbm6}
\fi
\ifamsf
  \newcommand{\Bbb}[1]{
      \mathchoice{\mbox{\bbbfont #1}}{\mbox{\bbbfont #1}}
      {\mbox{\smallbbbfont #1}}{\mbox{\tinybbbfont #1}}}
\else \def\Bbb{\bf} \fi
\iffrakf
  \newcommand{\frak}[1]{
      \mathchoice{\mbox{\frakfont #1}}{\mbox{\frakfont #1}}
      {\mbox{\smallfrakfont #1}}{\mbox{\tinyfrakfont #1}}}
\else \def\frak{\bf}
\fi

\catcode`\@=11
\long\def\@makefntext#1{
\protect\noindent \hbox to 3.2pt {\hskip-.9pt
$^{{\ninerm\@thefnmark}}$\hfil}#1\hfill}                

\def\@makefnmark{\hbox to 0pt{$^{\@thefnmark}$\hss}}  

\def\ps@myheadings{\let\@mkboth\@gobbletwo
\def\@oddhead{\hbox{}
\rightmark\hfil\ninerm\thepage}
\def\@oddfoot{}\def\@evenhead{\ninerm\thepage\hfil
\leftmark\hbox{}}\def\@evenfoot{}
\def\sectionmark##1{}\def\subsectionmark##1{}}

\renewcommand{\thefootnote}{\fnsymbol{footnote}}

\newcounter{sectionc}\newcounter{subsectionc}\newcounter{subsubsectionc}
\renewcommand{\section}[1] {\vspace*{0.6cm}\addtocounter{sectionc}{1}
\setcounter{subsectionc}{0}\setcounter{subsubsectionc}{0}\noindent
        {\normalsize\bf\thesectionc. #1}\par\vspace*{0.4cm}}
\renewcommand{\subsection}[1] {\vspace*{0.6cm}\addtocounter{subsectionc}{1}
        \setcounter{subsubsectionc}{0}\noindent
        {\normalsize\it\thesectionc.\thesubsectionc. #1}\par\vspace*{0.4cm}}
\renewcommand{\subsubsection}[1]
{\vspace*{0.6cm}\addtocounter{subsubsectionc}{1}
        \noindent
{\normalsize\rm\thesectionc.\thesubsectionc.\thesubsubsectionc.
        #1}\par\vspace*{0.4cm}}

\newcounter{appendixc}
\newcounter{subappendixc}[appendixc]
\newcounter{subsubappendixc}[subappendixc]

\renewcommand{\appendix}[1] {\vspace*{0.6cm}
        \refstepcounter{appendixc}
        \setcounter{figure}{0}
        \setcounter{table}{0}
        \setcounter{equation}{0}
        \renewcommand{\thefigure}{\Alph{appendixc}.\arabic{figure}}
        \renewcommand{\thetable}{\Alph{appendixc}.\arabic{table}}
        \renewcommand{\theappendixc}{\Alph{appendixc}}
        \renewcommand{\theequation}{\Alph{appendixc}.\arabic{equation}}
        \noindent{\bf Appendix \theappendixc #1}\par\vspace*{0.4cm}}

\def\abstracts#1{{

\centering{\begin{minipage}{12.2truecm}\footnotesize\baselineskip=12pt\noindent
        \centerline{\footnotesize ABSTRACT}\vspace*{0.3cm}
        \parindent=0pt #1
        \end{minipage}}\par}}

\renewenvironment{thebibliography}[1]
        {\begin{list}{\arabic{enumi}.}
        {\usecounter{enumi}\setlength{\parsep}{0pt}
\setlength{\leftmargin 1.25cm}{\rightmargin 0pt}
         \setlength{\itemsep}{0pt} \settowidth
        {\labelwidth}{#1.}\sloppy}}{\end{list}}

\newcounter{itemlistc}
\newcounter{romanlistc}
\newcounter{alphlistc}
\newcounter{arabiclistc}

\newcommand{\fcaption}[1]{
        \refstepcounter{figure}
        \setbox\@tempboxa = \hbox{\footnotesize Fig.~\thefigure. #1}
        \ifdim \wd\@tempboxa > 6in
           {\begin{center}
        \parbox{6in}{\footnotesize\baselineskip=12pt Fig.~\thefigure. #1}
            \end{center}}
        \else
             {\begin{center}
             {\footnotesize Fig.~\thefigure. #1}
              \end{center}}
        \fi}

\newcommand{\tcaption}[1]{
        \refstepcounter{table}
        \setbox\@tempboxa = \hbox{\footnotesize Table~\thetable. #1}
        \ifdim \wd\@tempboxa > 6in
           {\begin{center}
        \parbox{6in}{\footnotesize\baselineskip=12pt Table~\thetable. #1}
            \end{center}}
        \else
             {\begin{center}
             {\footnotesize Table~\thetable. #1}
              \end{center}}
        \fi}

\def\@citex[#1]#2{\if@filesw\immediate\write\@auxout
        {\string\citation{#2}}\fi
\def\@citea{}\@cite{\@for\@citeb:=#2\do
        {\@citea\def\@citea{,}\@ifundefined
        {b@\@citeb}{{\bf ?}\@warning
        {Citation `\@citeb' on page \thepage \space undefined}}
        {\csname b@\@citeb\endcsname}}}{#1}}

\newif\if@cghi
\def\cite{\@cghitrue\@ifnextchar [{\@tempswatrue
        \@citex}{\@tempswafalse\@citex[]}}
\def\citelow{\@cghifalse\@ifnextchar [{\@tempswatrue
        \@citex}{\@tempswafalse\@citex[]}}
\def\@cite#1#2{{$\null^{#1}$\if@tempswa\typeout
        {IJCGA warning: optional citation argument
        ignored: `#2'} \fi}}

 1
 1
 1

\font\ninerm=cmr9

\makeatother

\makeatletter
\def\@cite#1#2{{\if@cghi$\null^{#1}$%
\else
[#1]\fi\if@tempswa\typeout
        {IJCGA warning: optional citation argument
        ignored: `#2'} \fi}}
\makeatother

\newcommand{\eqref}[1]{Eq.~(\ref{#1})}
\newcommand{\eqsref}[1]{Eqs.~(\ref{#1})}
\newcommand{\U}{U}
\newcommand{\rowlattice}{\mathop{\rm rowlattice}\nolimits}
\newcommand{\trace}{\mathop{\rm trace}\nolimits}
\newcommand{\rank}{\mathop{\rm rank}\nolimits}
\newcommand{\Ph}{{\widehat\Phi}}
\newcommand{\Vh}{{\widehat{V}}}
\newcommand{\Sh}{{\widehat{\Sigma}}}
\newcommand{\eff}{\rm eff}
\newcommand{\fact}{{{1\over 2\pi\sqrt 2}}}
\newcommand{\factt}{{{i\over \sqrt 2}}}
\newcommand{\thetabar}{{\bar\theta}}
\newcommand{\transpose}{{\rm T}}

\begin{document}

\hbox to\hsize{\hfill \vbox{\baselineskip12pt\hbox{CLNS-95/1356}
\hbox{IASSNS-HEP-95/64}\hbox{hep-th/9508107}}}
\vspace*{0.4cm}
\centerline{\normalsize\bf TOWARDS MIRROR SYMMETRY AS DUALITY FOR}
\baselineskip=19pt
\centerline{\normalsize\bf TWO DIMENSIONAL ABELIAN GAUGE THEORIES%
\footnote{Talk given by M.R.P. at {\it Strings '95}.}}
\baselineskip=16pt

\vspace*{0.6cm}
\centerline{\footnotesize DAVID R. MORRISON\footnote{On
leave from: Department of Mathematics, Duke University,
Durham, NC 27708-0320 USA}}
\baselineskip=13pt
\centerline{\footnotesize\it Department of Mathematics, Cornell University,
Ithaca, NY 14853 USA}
\vspace*{0.3cm}
\centerline{\footnotesize and}
\vspace*{0.3cm}
\centerline{\footnotesize M. RONEN PLESSER\footnote{Permanent
 address (from 1 Aug.\ 1995): Department of Particle Physics,
Weizmann Institute of Science, Rehovot 76100, Israel}}
\baselineskip=12pt
\centerline{\footnotesize\it
School of Natural Sciences, Institute for Advanced Study,
Princeton, NJ 08540 USA}

\vspace*{0.9cm}
\abstracts{Superconformal sigma models with Calabi--Yau target spaces
described as complete intersection subvarieties in toric varieties can
be obtained as the low-energy limit of certain abelian gauge theories
in two dimensions.  We formulate mirror symmetry for this class of
Calabi--Yau spaces as a duality in the abelian gauge theory, giving
the explicit mapping relating the two Lagrangians.  The duality relates
inequivalent theories which lead to isomorphic theories in the
low-energy limit.  This formulation suggests that mirror symmetry could
be derived using abelian duality.  The application of duality
in this context is complicated by the presence of nontrivial dynamics
and the absence of a global symmetry.  We propose a way to overcome
these obstacles, leading to a more symmetric Lagrangian.  The argument,
however, fails to produce a derivation of the conjecture.
}


\normalsize\baselineskip=15pt
\setcounter{footnote}{0}
\renewcommand{\thefootnote}{\alph{footnote}}

\vspace*{0.6cm}
\noindent{\normalsize\bf Introduction}
\vspace*{0.4cm}

Two dimensional conformal field theories with $N{=}2$ supersymmetry
have been extensively studied as candidate vacua for perturbative
string theory.  A particularly interesting class of these consists of
supersymmetric sigma models with Calabi--Yau target spaces.  These
models exhibit a remarkable duality---known as mirror
symmetry\cite{dixon,lvw,cls,gp}---which relates two topologically distinct
target spaces leading to isomorphic conformal field theories.  The
duality has the property that classical computations in one model
reproduce
exact computations---including nonperturbative corrections---in the other.
This
property has been used to study both the geometry of Calabi--Yau
spaces and the properties of the CFT's they determine, with great
success.  However, a deep understanding of why such a duality exists has
been lacking.

The initial observations of mirror pairs in a
restricted class of Calabi--Yau spaces have been generalized in a set
of elegant conjectures by
Batyrev and
Borisov\cite{batyrev:mirror,borisov,batbor}.
These authors proposed a construction of the
mirror partner to a given Calabi--Yau space in a rather broad class
(the class of ``complete intersections in toric varieties''),
and offered some evidence that the space constructed was
indeed the mirror.  At roughly the same time, Witten\cite{phases}
noted that for precisely this class of
Calabi--Yau spaces, the associated conformal field theory
could be obtained as the low-energy approximation to a
supersymmetric abelian gauge theory in two
dimensions.  The coincidence of the two classes suggests the existence
of a natural interpretation of the conjectures of Batyrev and Borisov
in the context of Witten's model.  In this note we present this
interpretation, quantifying and making precise some earlier remarks
by a number of authors
to the effect that mirror symmetry must be
a manifestation of electric-magnetic duality
in these models.  We also present an attempt to establish the
equivalence of the two dual models by a modified version of abelian
duality, and show how the (present) attempt fails.

\section{The Gauged Linear Sigma Model}

We begin with a brief review of the gauged linear sigma model
(GLSM) construction of Witten\cite{phases}.
The model is formulated in (2,2)
superspace, and requires for its construction the specification
of a compact abelian group $G$, a faithful representation  $\rho:G\to\U(1)^n$,
and a  $G$-invariant polynomial $W(x_1,\dots,x_n)$,
where $x_1, \dots, x_n$ are coordinates in a
complex vector space ${\Bbb C}^n$
on which $\U(1)^n$ acts diagonally.
To construct the gauged linear sigma model, we begin with $n$ chiral
superfields $\Phi_i$ (satisfying $\overline{D}_+\Phi_i =
\overline{D}_-\Phi_i=0$) interacting via the holomorphic superpotential
$W(\Phi_1,\ldots,\Phi_n)$.  The model is invariant under the action of
$G$ (via $\rho$) on $\vec\Phi$ and we gauge this action, preserving $N{=}2$
supersymmetry, by introducing the ${\frak g}$-valued vector multiplet
$V$ with invariant field strength
$\Sigma=\frac1{\sqrt{2}}\overline{D}_+D_-V$.  This last field is
{\it twisted chiral}, which means that $\overline{D}_+\Sigma
=D_-\Sigma=0$.  For each continuous $\U(1)$ factor of $G$ we include
a Fayet-Illiopoulos $D$-term and a $\theta$-angle; these terms are
naturally written in terms of the complex combination  $\tau=ir+\frac
1{2\pi}\theta$.
The resulting Lagrangian density is thus
\begin{eqnarray}
{\cal L}&=&\int d^4\theta\left(\|e^{R({V})}\vec{\Phi}\|^2
-\frac1{4e^2}\|{\Sigma}\|^2\right)\nonumber\\
&&+ \left(\int d\theta^+d\theta^- W(\Phi_1,\dots,\Phi_n) +
 \mbox{ c.c. } \right)\label{GLSM}\\
&&+ \left( {i\over\sqrt 2}\,
\int d\theta^+d\bar\theta^- \langle\tau,\Sigma\rangle
+ \mbox{ c.c. } \right) ,\nonumber
\end{eqnarray}
where $R=-i\,d\rho:{\frak g}\to{\Bbb R}^n$ is the
derivative of the representation $\rho$ (with a factor of $-i$
 to make it real-valued).

Concretely, the general compact abelian group takes the form $G =
\U(1)^{n-d}\times\Gamma$ where $\Gamma$ is a product of finite cyclic groups.
Choosing a basis for the continuous part we have $n{-}d$ vector
multiplets $V_a$; the action on the fields is given by $R(V_a)\Phi_i =
Q_i^a\Phi_i$ for some integer charges $Q_i^a$.
Notice that the discrete group $\Gamma$ does not appear explicitly
in the Lagrangian, but it does affect the construction of the field
theory---the fields are sections of bundles
with structure group $G$.
In the case that $\Gamma$ is nontrivial,
we recover in this way an orbifold of
another theory (cf.\ Ref.~\citelow{phases}).

We will be interested in families of such models, parameterized by the
coefficients of $W$ and by the instanton factors $q_a := e^{2\pi
i\tau_a}$. (There will be a set of complex codimension one in the
parameter space along which the model is singular; we will study values
of the parameters away from this locus.)
A family is thus characterized by the group $G$ and the
collection of monomials appearing in $W$.
In order to specify these data, it is convenient
to introduce a $u\times n$ matrix $P$ of rank $d$ with nonnegative integer
entries, and a factorization $P=ST$ of $P$ as a product of integer matrices
$S$ and $T$, each of rank $d$.  The rows $t_\alpha$ of $T$ can then be
used to construct a collection of Laurent monomials
$x^{t_\alpha}:=\prod x_i^{t_{\alpha i}}$, and the group $G$ is defined to
be the largest subgroup of $H=\U(1)^n$ which leaves the monomials
$x^{t_\alpha}$
invariant.  The monomials $x^{p_r}$ defined by the rows of $P$ are then
$G$-invariant by construction, thanks to the relation
 $p_{ri}=\sum s_{r\alpha}t_{\alpha i}$.
Since $p_{ri}\ge0$ by assumption, we may use these monomials to specify
the family of interaction polynomials
\begin{equation}\label{W}
W(x_1,\dots,x_n):=\fact\sum_{r=1}^u c_r x^{p_r}=\fact\sum_{r=1}^u c_r
\prod_{i=1}^n x_i^{p_{ri}}\ .
\end{equation}
Alternatively, if we are given $G$ and a family of polynomials
$W$, it is not difficult to reconstruct
the matrices $P$, $S$, and $T$.  (Actually, $S$ and $T$ are only well-defined
up to $(S,T)\mapsto (SL,L^{-1}T)$, with $L$ an invertible integer matrix.)

If we start with only $W$, as specified by the rows of the matrix
$P$, then the choice of a factorization $P=ST$ amounts to a choice of a
subgroup $G\subset \U(1)^n$ (which can be less than maximal) under which $W$
is invariant.
On the other hand, if we start with only $G$ then the choice of
factorized matrix
$P=ST$ allows us to specify which of the $G$-invariant monomials
should be included in $W$.  In particular, it is possible to omit
some monomials and in this way to study
subsets of the
maximal set of all $G$-invariant superpotentials. As we shall see,
this possibility is useful.  (The assumption made above on the rank of
$S$ and $T$ restricts this choice---we must use ``enough'' monomials
to get the rank to be correct---but the restriction is not
essential and relaxing it would simply require a more cumbersome notation.)

Under these conditions, the
possible choices for factorizations (and hence for $G$) for a given $P$
are rather limited.  In any factorization $P=ST$, the rows of $T$ will
generate an integral lattice of rank $d$, and the rows of $P$ will
generate a sublattice, also of rank $d$.  This implies that the quotient
\[\rowlattice(T)/\rowlattice(P)\]
is a torsion subgroup of ${\Bbb Z}^n/\rowlattice(P)$. There are only
finitely many of choices of such a torsion subgroup, once $P$
has been fixed.

\section{The $R$-Symmetry}

The gauged linear sigma model is not conformally invariant, and will
flow to strong-coupling in the infrared.  Generically such a theory
would be expected to develop a mass gap, but
evidence has been found\cite{phases,SilWit} that the theory at
the IR fixed point will be nontrivial if we require that
the high-energy theory admit a non-anomalous chiral $R$-symmetry.
We consider a right-moving $R$-symmetry, acting in such a way that
 $\theta^+$ has charge $1$ and $\theta^-$ is neutral.
Invariance of the kinetic terms requires that
the gauge fields ${V}$ be neutral, and their field strengths
${\Sigma}$ hence have charge $1$.  Invariance of the superpotential
interaction requires that the superpotential have charge $1$.  If we let
$\mu_i$ denote the $R$-charge of $\Phi_i$ (which may be a rational number)
this tells us that
\[\sum_{i=1}^n  p_{ri} \mu_i=1 \quad \mbox{for all $r$}\ .\]

This chiral symmetry can be anomalous in the presence of the gauge
fields. A quick computation\cite{phases,summing} shows that
the anomaly is given by a function on the Lie
algebra proportional to $V\mapsto\trace(R(V))$;
we require that this vanish identically, i.e., that the symmetry be
nonanomalous. Since the action of the continuous part
of $G$ on the monomial $x_1\cdots x_n$ is via
$\exp(\trace(R(V)))$, this is the same as requiring
that $x_1\cdots x_n$
be invariant under the continuous part of $G$ (or in our explicit
coordinates that $\sum_i Q_i^a=0$ for all $a$).  This in turn will hold
exactly when the vector of exponents $(1,\dots,1)$ is a linear combination
of the rows of the matrix $T$, with rational coefficients.  (If we had wanted
$x_1\cdots x_n$ to be invariant under all of $G$, we would have insisted
upon integer coefficients.)  That is, there is a rational
vector $\vec{\lambda}$ such that $\vec{\lambda}^\transpose T=(1,\dots,1)$.

Since we are assuming that the $d\times u$ matrix $S^\transpose$
has rank $d$, it
is possible to solve the equation $S^\transpose\vec{\nu}=\vec{\lambda}$ for
a rational vector $\vec{\nu}$. If we also set
$\vec{\mu}=(\mu_1,\dots,\mu_n)^\transpose$, then
 we can write the conditions for an
unbroken $R$-symmetry as the existence of $\vec\mu,\,\vec\nu$ such
that
\begin{eqnarray}
P\vec{\mu}&=&(1,\dots,1)^\transpose\label{eq1}\\
\vec{\nu}^\transpose P&=&(1,\dots,1)\ .\label{eq2}
\end{eqnarray}

Finally, using the $R$-symmetry and
calculating as in Ref.~\citelow{SilWit}, one finds that
the central charge $c$ of the fixed-point CFT is determined by
\[d-(c/3)=2\sum_{i=1}^n\mu_i=2\,\vec{\nu}\,{}^\transpose
P\vec{\mu}=2\sum_{r=1}^u\nu_r\ .
\]

It may be useful to have some examples.

\vspace*{0.6cm}\noindent{\normalsize\it Example 1.}\par\vspace*{0.4cm}

Consider $1\times1$ matrices
$(p)=(s)(t)$ with $s$ and $t$ positive integers.  Note that
\eqsref{eq1} and (\ref{eq2}) are trivially satisfied in this case. Since
$G$ is the group
which leaves $x^t$ invariant, we must have $G={\Bbb Z}/(t)$.  The interaction
polynomial is $W(x)=x^p$, so our model is an orbifold of the
Landau-Ginsburg version of a minimal model.

\vspace*{0.6cm}\noindent{\normalsize\it Example 2.}\par\vspace*{0.4cm}

Consider a $u\times 6$ matrix $P$ of rank $5$ such
that $p_{0i}=p_{r0}=1$, (letting the $i$ and $r$ indices start at $0$
for this example) implementing \eqsref{eq1} and (\ref{eq2}) directly.
This leads to a polynomial of the form
\[W(x_0,\dots,x_5)=x_0\,W'(x_1,\dots,x_5)=
\fact\left(c_0\,x_0x_1x_2x_3x_4x_5+x_0\,f(x_1,\dots,x_5)\right),\]
where $f$ is a polynomial in $5$ variables containing precisely
$u{-}1$ monomials.  The existence of a kernel for $p$, an integral
generator of which we denote by $(-k,\ell_1,\dots,\ell_5)^\transpose$,
means that $W$ is invariant under a $\U(1)$-action on
$(x_0,\dots,x_5)$ with weights
$(-k,\ell_1,\dots,\ell_5)$---this implies that $f$ is
quasi-homogeneous of degree $k$ and that $k=\sum \ell_i$.
One possible factorization $P=ST$
would then be given by taking the rows of $T$ to be a basis for the
$\U(1)$-invariant Laurent monomials, and expressing the monomials
appearing in $f$ in terms of those.  By construction, this leads to
$G=\U(1)$.  In general (if $f$ is a generic quasi-homogeneous
polynomial) this is the only possible factorization.  For special
subfamilies (determined by a subset of the rows of
the maximal $P$) the
polynomial is invariant under additional discrete symmetries (the
restriction on $\rank(P)$ constrains the continuous part), and
other factorizations are possible, leading
to groups $G$ with a nontrivial discrete part.

As we will see below, the weights we have given specify a weighted
projective space ${\Bbb P}^{(\ell_1,\dots,\ell_5)}$, and
$\{f{=}0\}\subset {\Bbb P}^{(\ell_1,\dots,\ell_5)}$
defines a (singular) Calabi--Yau hypersurface in that space,
closely related to the GLSM built from the data $P=ST$.
A particularly interesting case is when $u=6$, and $P$ is $6\times6$.
The description of these models in terms of the matrix $P$ was
first given by Candelas, de la Ossa, and Katz\cite{cdk} in the
course of generalizing the Berglund--H\"ubsch\cite{BHub} construction.

\vspace*{0.6cm}\noindent{\normalsize\it Example 3.}\par\vspace*{0.4cm}

To make the previous example a bit more concrete, consider a Fermat-type
polynomial
\[f(x_1,\dots,x_5)=x_1^{a_1}+x_2^{a_2}+x_3^{a_3}+x_4^{a_4}
+x_5^{a_5}\]
with $\ell_j=k/a_j$ and suppose that $\ell_5=1$.
Choosing $G=\U(1)$ as in example 2 corresponds to the factorization
\[
\left(\begin{array}{cccccc}
1&1&1&1&1&1\\ 1&a_1&0&0&0&0\\ 1&0&a_2&0&0&0\\ 1&0&0&a_3&0&0\\
1&0&0&0&a_4&0\\ 1&0&0&0&0&a_5
\end{array}\right)=
\left(\begin{array}{ccccc}
1&1&1&1&1\\ 1&a_1&0&0&0\\ 1&0&a_2&0&0\\ 1&0&0&a_3&0\\ 1&0&0&0&a_4\\
1&0&0&0&0
\end{array}\right)
\left(\begin{array}{cccccc}
1&0&0&0&0&k\\ 0&1&0&0&0&-\ell_1\\ 0&0&1&0&0&-\ell_2\\
0&0&0&1&0&-\ell_3\\ 0&0&0&0&1&-\ell_4
\end{array}\right).
\]
On the other hand, to maximize the group $G$ we should use the entire
row space of $P$ as the row space of $T$ which leads to the factorization
\[
\left(\begin{array}{cccccc}
1&1&1&1&1&1\\ 1&a_1&0&0&0&0\\ 1&0&a_2&0&0&0\\ 1&0&0&a_3&0&0\\
1&0&0&0&a_4&0\\ 1&0&0&0&0&a_5
\end{array}\right)=
\left(\begin{array}{ccccc}
1&0&0&0&0\\ 0&1&0&0&0\\ 0&0&1&0&0\\ 0&0&0&1&0\\ 0&0&0&0&1\\
k&-\ell_1&-\ell_2&-\ell_3&-\ell_4
\end{array}\right)
\left(\begin{array}{cccccc}
1&1&1&1&1&1\\ 1&a_1&0&0&0&0\\ 1&0&a_2&0&0&0\\ 1&0&0&a_3&0&0\\
1&0&0&0&a_4&0
\end{array}\right).
\]
This corresponds to an orbifold of the previous model; in fact, this
is the quotient found by Greene and Plesser\cite{gp} to correspond to
the mirror manifold.  Thus we can reproduce the construction of
Ref.~\citelow{gp} in this context by appropriate choices of factorization.

\section{The Low-Energy Limit}

As explained in detail in Ref.~\citelow{phases}, the low-energy limit of this
theory can be described explicitly and---when the central charge is
an integer---often coincides with a sigma-model
on a Calabi--Yau space (for appropriate values of the parameters).

To study the low-energy limit one begins by mapping out the space of
classical vacua of the theory.  To this end, we first solve the algebraic
equations of motion for the auxiliary fields $D_a$ (in the vector
multiplets) and $F_i$ (in the chiral multiplets)
\begin{eqnarray}
D_a &=& -e^2\left(\sum_{i=1}^n Q_i^a |\phi_i|^2 - r_a\right)\label{D}\\
F_i &=& -{\partial W\over\partial\phi_i}\ .\label{F}
\end{eqnarray}
The potential energy for the bosonic zero modes is then
\[U = {1\over 2e^2}\sum_{a=1}^{n-d}D_a^2 + \sum_{i=1}^n |F_i|^2 +
\sum_{a,b}\bar\sigma_a\sigma_b\sum_{i=1}^n Q_i^aQ_i^b|\phi_i|^2\ ,\]
where $\phi_i,\,\sigma_a$ are the lowest components of
$\Phi_i,\,\Sigma_a$ respectively.  The space of classical vacua is the
quotient by $G$ of the set of zeros of $U$.

Neglecting for a moment the superpotential, the space of solutions
(setting $\sigma=0$) is $D^{-1}(0)/G$.  For values of the instanton
factors $q_a$ in a suitable range this is a toric variety of dimension
$d$. In the ``geometric'' case in which $\vec{\mu}$ has $N$ $1$'s and $n{-}N$
$0$'s, this variety can be recognized as the total space of a sum of
$N$ line bundles over a {\it compact}\/ toric variety of dimension $d-N$.
The equations $F_i=0$ then determine (for generic values of the
parameters $c_r$) a complete intersection subvariety of codimension
$N$ in the base space, homologous to the intersection of the divisors
associated to the $N$ line bundles.  The condition in \eqref{eq2}
implies that this subvariety is Calabi--Yau.

The fixed-point CFT is given by the nonlinear sigma model with this
target space.  The moduli of this CFT are the marginal operators in
\eqref{GLSM}. Na\"{\i}vely, both the $q_a$ and the $c_r$ would appear to
be marginal
couplings.  The latter, however, are not all independent. As is
well known, some of them can be absorbed in rescalings of the fields
$\Phi_i$. The true moduli are thus the $q_a$ and the scaling-invariant
combinations of the $c_r$. There will in general be other marginal
operators in the model which do not appear explicitly in the
Lagrangian; we restrict our attention to the subspace of those that do.
Of course, the linear (or more accurately, toric) structure with which we
have endowed our moduli space is an artifact of our description.  In
particular, it is consistent with the natural complex structure that
this moduli space carries intrinsically, but bears no relation to the
K\"ahler structure determined by the Zamolodchikov metric.  In other
words, the $q_a$ and $c_r$ are not the ``special'' coordinates on this
space.

For example, in example 2 above in which $G=\U(1)$, the $D$-term equation is
\[-k\,|x_0|^2+\sum_{i>0}\ell_i|x_i|^2 - r=0,\]
so when $r>0$ we cannot have $x_i=0$ for all $i>0$.  The space of
solutions $D^{-1}(0)/G$ can be recognized as the total space of the
canonical bundle over ${\Bbb P}^{(\ell_1,\dots,\ell_5)}$.  If we impose
the $F$-term equation as well
\[\sum_{i=0}^5\left|\frac{\partial W}{\partial x_i}\right|^2
=|W'|^2+|x_0|^2\sum_{i=1}^5\left|\frac{\partial W'}{\partial x_i}\right|^2
=0,\]
then for a generic choice of the coefficients of $W$ (away from a
codimension-one subspace we avoid as promised above) the space of
solutions is given by $x_0=0$ and $W'(x)=0$, yielding a
Calabi--Yau  hypersurface in ${\Bbb P}^{(\ell_1,\dots,\ell_5)}$.  Further
study shows that the low-energy excitations are tangent to this, so
that the low-energy theory is a nonlinear sigma model with this Calabi--Yau
target space.

\section{Mirror Symmetry}

We are now in a position to state the mirror symmetry conjectures for
this class of models.  Mirror symmetry relates two Calabi--Yau
manifolds which lead to isomorphic CFT's when used as target spaces
for supersymmetric nonlinear sigma models. The mirror isomorphism
reverses the sign of the right-moving
$R$-symmetry; many of the fascinating properties of mirror pairs can
be traced to this feature. Given a GLSM family determined by $P=ST$ we
will construct the mirror family in the same class of models.
We incorporate the sign change on $R$ by writing the
dual model using {\it twisted}\/ chiral matter fields coupled to twisted
gauge multiplets (with {\it chiral}\/ field strengths). We will call such
a model a {\it twisted}\/ gauged linear sigma model. As above, we
characterize the family of models by a factorized matrix of
nonnegative integers $\widehat P = \widehat S\widehat T$.  We use this
data to construct a twisted superpotential $\widehat W$ and write a
Lagrangian density (compare \eqref{GLSM})
\begin{eqnarray}
{\cal L}&=&\int d^4\theta\left(-\|e^{\widehat R({\widehat V})}
\widehat {\vec{\Phi}}\|^2
+\frac1{4e^2}\|{\widehat \Sigma}\|^2\right)\nonumber\\
&&+ \left(\int d\theta^+d\bar\theta^- \widehat W(\Ph) +
 \mbox{ c.c. } \right)\label{TGLSM}\\
&&+ \left( {i\over\sqrt 2}\,
\int d\theta^+d\theta^- \langle\widehat \tau,\widehat \Sigma\rangle
+ \mbox{ c.c. } \right) ,\nonumber
\end{eqnarray}
where $\widehat\Sigma = -{1\over 2\sqrt 2}\bar D_+\bar D_-\widehat V$ is the
(chiral) field strength.

The {\it mirror conjecture}\/ states that if we set
\begin{eqnarray}
\widehat P &=& P^\transpose\nonumber\\
\widehat S &=& T^\transpose\label{mirp}\\
\widehat T &=& S^\transpose\nonumber
\end{eqnarray}
the two families of CFT's defined by the infrared limits of the two
linear models are isomorphic.  This is essentially just a translation
of the conjecture of Batyrev and Borisov\cite{batbor}, versions of
which have appeared in Refs.~\citelow{ag:rigid,cdk}.
In fact, the statement given here generalizes the
conjecture somewhat since the factorization makes possible a map
between subfamilies.

One can make a more precise statement of the conjecture---specifying
explicitly the map between the two parameter spaces which relates
isomorphic low-energy limits
\begin{eqnarray}
q_a &=& \prod_{r=1}^u \widehat c_i^{Q_i^a}\nonumber\\
\widehat q_{\widehat a} &=& \prod_{i=1}^n c_r^{\widehat Q_r^{\widehat
a}}\ .\label{mondiv}
\end{eqnarray}
Note that as expected only the true moduli appear on the right-hand
side of \eqref{mondiv}.
The asymptotic limits of \eqref{mondiv} constitute
the {\it
monomial-divisor mirror map}, first proposed in Refs.\ \citelow{catp,mondiv}
(see also Ref.~\citelow{Bat:qu})
and extended in Refs.\ \citelow{ag:rigid,multdiv}.
The ability to
write a global version of this
map and not just an asymptotic form is related to the
coordinates we use on the moduli space (for a full discussion of this
including a proof of \eqref{mondiv} for a class of examples see
Ref.~\citelow{summing}).

We note that this statement of the conjecture is very reminiscent of
recent results on duality in four-dimensional supersymmetric gauge
theories\cite{SWI,SWII,S}. Two distinct gauge theories
with nontrivial dynamics lead to the same low-energy physics; further,
instanton effects in one model are reproduced by classical physics
in the other.

In example 3 above, it should be clear that this construction
interchanges the two choices of factorization, leaving the (symmetric)
matrix $P$ unchanged.  As mentioned above, this reproduces the
orbifold construction of the mirror in Ref.~\citelow{gp}, clearly and
explicitly demonstrating how the general construction reduces to this
upon restricting to a subfamily of all possible $W$---precisely the
subfamily invariant under the maximal discrete group.

\section{Abelian Duality}

The formulation we have given for the mirror conjecture in the context of
the GLSM immediately suggests a relation to abelian duality.  We recall
that given a theory with an abelian continuous global symmetry, one can
use duality to obtain an equivalent theory. The dual model is obtained
by gauging the global symmetry, introducing Lagrange multipliers which
constrain the connection to be pure gauge (ensuring that the theory is actually
unchanged).  One must then eliminate the original degrees of freedom,
in the process generating a nontrivial effective
action for the Lagrange multipliers, which become the fundamental degrees
of freedom for the dual model.  Classically this is accomplished by
gauge-fixing.  In a theory with
$N{=}2$ supersymmetry and chiral charged fields, the Lagrange multipliers
mentioned above will be {\it twisted\/} chiral fields, and will appear
in the twisted superpotential as $\widehat\Lambda\Sigma$.  Thus the dual
variables will naturally be twisted chiral\cite{GHR}, as expected for
the mirror model. This idea was pursued in Refs.\ \citelow{G,RV,GW,BH}, and in
special cases leads to an interpretation of mirror symmetry as abelian
duality.
In the general case, however, the approach runs into difficulties.  The
most obvious one is that a generic Calabi--Yau  manifold $M$ has no isometries,
hence the associated two dimensional CFT lacks suitable global symmetries.

If this approach is to lead to the equivalence of GLSM models which is
tantamount to the mirror conjecture, some modification will be required.
The two linear models are {\it not\/}
equivalent; the conjecture implies only that they become equivalent
in the extreme low-energy limit.  This is consistent with the fact
that the model has nontrivial dynamics. As discussed above this is
similar to the recent discoveries in four-dimensional supersymmetric
gauge theories.

In the context of the GLSM, there is a natural symmetry
group one would attempt to use.  This is the group $H = \U(1)^n$ acting
by phases on the fields $\Phi_i$.  The difficulty, of course, is that
this symmetry is explicitly broken by the superpotential \eqref{W} to the
subgroup $G$. Since this symmetry is gauged in \eqref{GLSM}, the
resulting model has no global symmetries at all (except the $R$-symmetry
which we cannot gauge).\footnote{This once led E. Witten to
describe the problem of understanding mirror symmetry in this model as
the question ``How to perform duality on a non-symmetry?''} We can
attempt to overcome this obstacle by recasting the symmetry-breaking
terms as anomalies. To implement this idea, introduce a set of
twisted chiral fields $\widehat\Psi_r$ into the model, coupled to
twisted gauge multiplets $\widehat V_r$ gauging the
group $\widehat H = \U(1)^u$ which
acts by phases, i.e.\ consider the additional term in the
kinetic energy
\begin{equation}
{\cal L}_{\widehat k} = \int d^4\theta
\left( -\|\widehat{\vec\Psi} e^\Vh\|^2 +{1\over 4e^2}\|\Sh\|^2\right)\ .
\label{khat}
\end{equation}
We then replace the superpotential \eqref{W} by
\begin{equation}
W_s = \fact  \sum_{r=1}^u\Sh_r\left[\log\left(
c_r\prod_{i=1}^n\Phi_i^{p_{ri}}\right)+1\right]\ .
\label{Ws}
\end{equation}
In this form it is clear that $H$ is broken by $\widehat H$ anomalies, as
desired.  Note that this interaction does not break the $R$-symmetry.
However, the couplings $c_r$ have nonzero beta functions,
which will cause them to grow large, so the $\widehat H$ gauge theory is
strongly coupled at low energies.  In this limit, as discussed in
Refs.\ \citelow{phases,summing}, this sector of the model is in a confining
phase, in which the lowest component $\widehat\sigma$ of $\Sh$ gets
a nonzero expectation value.  The charged fields are then all massive,
with masses of order this expectation value, and the
light degrees of freedom are in the $\Sh$ multiplet.
The effective action for these can be reliably computed at
one-loop order.  Integrating out the $\widehat\Psi$'s leads to the effective
superpotential
\begin{equation}
W_{\eff} = \fact\sum_{r=1}^u \Sh_r\left[\log\left(
c_r\prod_{i=1}^n\Phi_i^{p_{ir}}\right)+1 - \log\Sh_r\right]\ .
\label{Wsef}
\end{equation}
In the effective theory we can treat $\Sh$ as a chiral field; since
there are no charged fields we can forget its relation to a gauge
symmetry. At low energy, we can eliminate $\Sh$ from \eqref{Wsef} to get
$\Sh_r = c_r\prod_{i=1}^n\Phi_i^{p_{ir}}$.  When substituted in
\eqref{Wsef}, this leads to \eqref{W} as the effective superpotential for
$\Phi$.  Thus at low energies this model is equivalent to the original
GLSM, while the symmetry-breaking terms are explicitly exhibited as
anomalies.

We now wish to perform a duality transformation with respect to the
anomalous symmetry.  Gauging an anomalous symmetry appears inconsistent,
but in performing a duality transformation an anomalous abelian
symmetry can be restored
by assigning transformation properties (under the twisted symmetries) to
the Lagrange multipliers.\cite{GR} In the case at hand, gauging
the global symmetry means that the gauge group becomes all of $H$.
We should also introduce Lagrange multipliers, transforming under
$\widehat H$, to constrain the field strength to lie in $\frak
g\subset\frak h$.  Classically, integrating these out will reproduce
the original model.  This is not true quantum mechanically, however,
as is most easily seen by introducing a small kinetic term for the
Lagrange multipliers, and considering the limit as this term is
removed.  It then becomes clear that the contribution of these to the
one-loop effective superpotential for $\Sh$ is constant in the limit.
Introducing these extra fields would thus destroy the one-loop
calculation that led to \eqref{Wsef}.

This suggests that we
consider a related Lagrangian, in which the twisted chiral matter is
an accurate ``reflection'' of the chiral matter.  This defines a model
that is manifestly mirror-symmetric.  Our proposal for this  new model
contains chiral
fields $\vec\Phi$ and twisted chiral fields $\widehat{\vec\Phi}$, and
(twisted) gauge multiplets gauging the entire group $H$ ($\widehat H$).
We begin with the Lagrangian density
\begin{eqnarray}
{\cal L} &=& \int d^4\theta
\left( \sum_i \left( J_i - {1\over 4e^2}|\Sigma_i|^2\right) -
\sum_r\left( \widehat J_r - {1\over 4e^2}|\Sh_r|^2\right) +
{1\over 2\pi}\sum_{ri}p_{ri} \log J_i \log \widehat J_r \right)\nonumber\\
&& + \left(\factt\int d\theta^+d\theta^-\langle\Sh,\widehat\tau\rangle
+{\rm c.c}\right)
+ \left(\factt\int d\theta^+d\thetabar^-\langle\Sigma,\tau\rangle
+{\rm c.c}\right) \ ,\label{our}
\end{eqnarray}
where $J_i = |\Phi_ie^{V_i}|^2$ and likewise for $\widehat J_r$.

This action is manifestly invariant under $H\times\widehat H$. It is
classically equivalent to the action we obtain by
integration by parts,
\begin{eqnarray}
{\cal L_{\rm eq}} &=& \int d^4\theta
\left( \sum_i \left( J_i - {1\over 4e^2}|\Sigma_i|^2\right) -
\sum_r\left( \widehat J_r - {1\over 4e^2}|\Sh_r|^2\right) +
{2\over \pi}\sum_{ri}p_{ri} V_i\widehat V_r\right)\nonumber\\
&& + \left(\int d\theta^+d\theta^- W(\Sh,\Phi)+{\rm c.c}\right)
+ \left(\int d\theta^+d\thetabar^- \widetilde W(\Sigma,\Ph)
+{\rm c.c}\right) \ ,\label{ourpi}
\end{eqnarray}
where the modified superpotentials are
\begin{eqnarray}
W(\Sh,\Phi) &=& \fact\sum_{r=1}^u \Sh_r\left[\log\left(
\widehat q_r\prod_{i=1}^n\Phi_i^{p_{ri}}\right)+1\right] \nonumber\\
\widetilde W(\Sigma,\Ph) &=& \fact\sum_{i=1}^n \Sigma_i\left[\log\left(
q_i\prod_{r=1}^u\Ph_r^{-p_{ri}}\right)+1\right]\ ,\label{Wt}
\end{eqnarray}
We see that the conditions for a
nonanomalous $R$-symmetry are precisely \eqsref{eq1} and (\ref{eq2}).
When these hold we expect to find at low energies an $N{=}2$
superconformal field theory where the $R$-symmetry defined above
becomes the chiral $\U(1)$ contained in the superconformal algebra.
We will use the second formulation as our definition of the quantum
theory. (In the presence of instantons the integration by parts
requires the consideration of boundary terms, which are nontrivial.)

It is worthwhile noting the way in
which \eqref{ourpi} manages to be
gauge-invariant despite the manifestly non-invariant interactions.
Under a gauge transformation (in $H\times\widehat H$) the variation of
\eqref{Wt} is cancelled, up to a total derivative, by the variation of
the $V\widehat V$ term.  Thus the full action is invariant under gauge
transformations approaching the identity at infinity, while
transformations with constant parameter are still anomalous.  We note
that this cancellation holds precisely when the exponents in the two
superpotentials are related as in \eqref{Wt}.  In a derivation of
mirror symmetry along these lines this would be the origin of the
first part of \eqref{mirp}.

At this point one still needs to show that the theory defined by
\eqref{ourpi} is equivalent to \eqref{GLSM}.  This can be achieved
using the fact that the coefficients $q_i,\,\widehat q_r$ once more
contain redundant deformations which can be undone by field rescalings
(in fact this is precisely the holomorphic extension of the statement
above about anomalies).  One can perform a redundant deformation of
the model to find a region in parameter space in which the $\Sh,\Ph$
system is in a confining phase and the one-loop approximation that led
to \eqref{Wsef} is valid.  One can also show that in this region the
integration over $\Ph$ will lead to constraints on
$\Sigma$, restricting the gauge group to $G$.  In fact, the last two
equations of \eqref{mirp} arise naturally in this context as well.

The manifest mirror
symmetry would then lead one to expect that \eqref{ourpi}  will also
be equivalent
to \eqref{TGLSM} under the conditions \eqsref{mirp} and (\ref{mondiv}).
To see this one performs redundant deformations to a region in
parameter space in which the $\Sigma,\Phi$ system confines and follows
a ``mirror'' version of the argument sketched above.   This naturally
incorporates all of the conditions in \eqsref{mirp} and (\ref{mondiv}).
We hasten to point out however that \eqref{ourpi} is not in fact
mirror symmetric.  The reason for
this is the sign in the second of \eqref{Wt}.\footnote{This sign was
inadvertently omitted in earlier stages of this work, leading to
conclusions---reported in several talks by the
authors---differing from those presented here.}
Unexpectedly, this
sign, which is required for gauge invariance since it follows by
integration by parts from \eqref{our}, cannot
be removed (it is related to the relative sign of the kinetic terms
for the chiral and twisted chiral fields).
Similar signs appear often in discussions of duality, and
are usually harmless.  In the case at hand, however, the sign
multiplies a logarithm, and its effect is to lead, at least
superficially, to a twisted superpotential polynomial in $\Ph^{-1}$
rather than $\Ph$ after performing the ``mirror'' argument.

This would seem to shatter the hopes for an understanding of mirror
symmetry for these models along the lines presented here.  However,
the model we present does seem to incorporate naturally many of the
features required for a derivation of the conjectures in section four.
It is possible that some modification of this model will lead to the
desired result.  Future study will tell if this is indeed the case.

\vspace*{0.6cm}
\noindent{\normalsize\bf Acknowledgements}
\vspace*{0.4cm}

We thank P. Candelas and X. de la Ossa for many discussions and for
collaboration in the early stages of this work, and
P. Argyres, P. Aspinwall, B. Greene, G. Moore, M. Ro\v cek,
and especially E. Witten and N. Seiberg for discussions.
The work of D.R.M.\ was supported in part by the United States Army
Research Office through the Mathematical Sciences Institute of
Cornell University, contract DAAL03-91-C-0027,
and in part by the National Science Foundation through
grants DMS-9401447 and PHY-9258582.
The work of M.R.P.\ was supported by  National Science Foundation grant
PHY-9245317 and by the W.M. Keck Foundation.

\vspace*{0.6cm}
\noindent{\normalsize\bf References}


\end{document}